\documentclass[letterpaper]{article} 
\usepackage{aaai2026}  
\usepackage{times}  
\usepackage{helvet}  
\usepackage{courier}  
\usepackage[hyphens]{url}  
\usepackage{graphicx} 
\usepackage{natbib}  
\usepackage{caption} 
\usepackage{algorithm}
\usepackage{algorithmic}

\usepackage{newfloat}
\usepackage{listings}

\usepackage{booktabs} 
\usepackage[table]{xcolor}
\usepackage{multirow}
\usepackage{colortbl}
\usepackage{subcaption}

\usepackage{amssymb}
\usepackage{amsmath}
\usepackage{pifont}
\usepackage{makecell}
\usepackage{adjustbox}
\usepackage{textcomp}

\usepackage{titlesec}

\DeclareCaptionStyle{ruled}{labelfont=normalfont,labelsep=colon,strut=off} 
\floatstyle{ruled}
\newfloat{listing}{tb}{lst}{}
\floatname{listing}{Listing}
\title{TraceTrans: Translation and Spatial Tracing for Surgical Prediction}
\author{
    Xiyu Luo\textsuperscript{\rm 1}\equalcontrib, Haodong Li\textsuperscript{\rm 1}\equalcontrib, Xinxing Cheng\textsuperscript{\rm 2}, He Zhao\textsuperscript{\rm 3}, Yang Hu\textsuperscript{\rm 4}, Xuan Song\textsuperscript{\rm 5}, Tianyang Zhang\textsuperscript{\rm 6}\thanks{Corresponding author.}\
}
\affiliations{
    \textsuperscript{\rm 1}Southern University of Science and Technology, China\\
    \hspace{1.5em}%
    \textsuperscript{\rm 2}University of Birmingham, United Kingdom\\
    \hspace{1.5em}%
    \textsuperscript{\rm 3}University of Liverpool, United Kingdom\\
    \hspace{1.5em}%
    \textsuperscript{\rm 4}University of Leicester, United Kingdom\\
    \hspace{1.5em}%
    \textsuperscript{\rm 5}Jilin University, China\\
    \hspace{1.5em}%
    \textsuperscript{\rm 6}University of Oxford, United Kingdom\\
    \hspace{1.5em}%

    luoxy2022@mail.sustech.edu.cn
    \hspace{2em}
    tianyang.zhang@eng.ox.ac.uk
}

\usepackage{bibentry}

\begin{document}

\maketitle

\begin{abstract}
Image-to-image translation models have achieved notable success in converting images across visual domains and are increasingly used for medical tasks such as predicting post-operative outcomes and modeling disease progression. However, most existing methods primarily aim to match the target distribution and often neglect spatial correspondences between the source and translated images. This limitation can lead to structural inconsistencies and hallucinations, undermining the reliability and interpretability of the predictions. These challenges are accentuated in clinical applications by the stringent requirement for anatomical accuracy. In this work, we present TraceTrans, a novel deformable image translation model designed for post-operative prediction that generates images aligned with the target distribution while explicitly revealing spatial correspondences with the pre-operative input. The framework employs an encoder for feature extraction and dual decoders for predicting spatial deformations and synthesizing the translated image. The predicted deformation field imposes spatial constraints on the generated output, ensuring anatomical consistency with the source. Extensive experiments on medical cosmetology and brain MRI datasets demonstrate that TraceTrans delivers accurate and interpretable post-operative predictions, highlighting its potential for reliable clinical deployment. 
\end{abstract}

\begin{links}
    \link{Code}{https://github.com/xyluoxy/TraceTrans}
\end{links}

\section{Introduction}
Medical image-to-image translation is particularly important for post-operative outcome prediction, where the objective is to generate realistic post-operative images from pre-operative inputs while preserving anatomical plausibility. Such predictions help clinicians visualize potential surgical outcomes, plan interventions, and communicate expected changes to patients. Beyond cosmetic surgery, similar approaches can model longitudinal changes in neurological diseases or tumor evolution, offering valuable insights for both clinical practice and research.

Recent advances in deep generative models, including GAN-based \cite{bib:pix2pix, bib:cyclegan} and diffusion-based approaches \cite{bib:palette, bib:latentDiffusion, bib:DDIB}, have improved image realism across translation tasks. However, most existing methods focus on matching the target distribution and neglect spatial correspondences between source and translated images, leading to inconsistencies and hallucinations that compromise reliability and interpretability. A common workaround is to enforce structural invariance between input and output to improve consistency \cite{zhang2019noise}, but this conflicts with the objective of post-operative prediction, where structural changes must be realistically modeled.

\begin{figure}[t!]
    \centering
    \includegraphics[width=\linewidth]{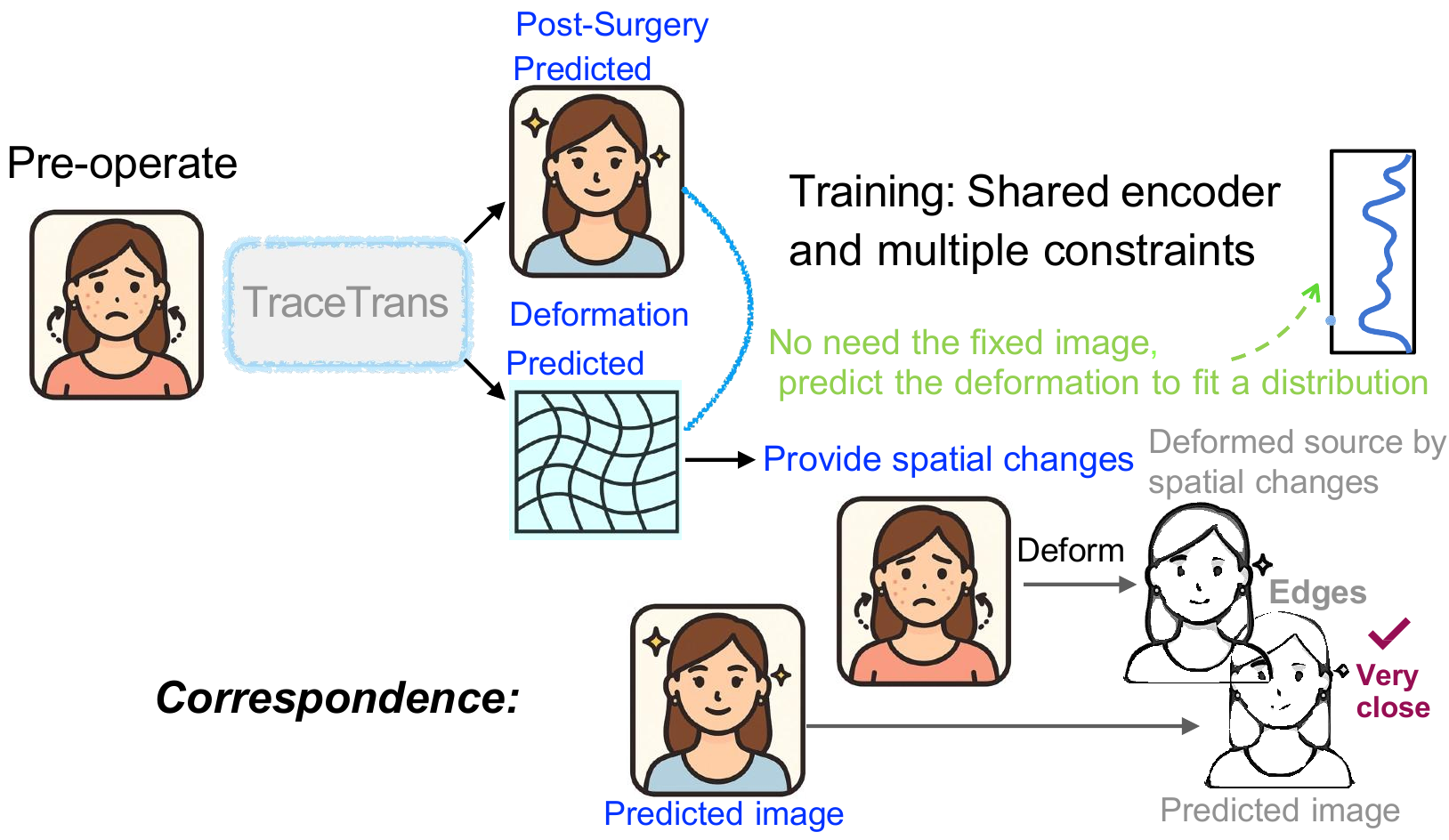}
    \caption{TraceTrans employs an encoder with dual decoders to predict both spatial deformation and translated images. The deformation field provides pixel-level correspondences from the source to the target, ensuring structural alignment while adapting to the target distribution. No fixed reference image is required during training and inference.}
    \label{fig:intro-tracetrans}
\end{figure}

To address the challenges of structural inconsistency and provide traceable spatial changes during translation for post-operative prediction, we aim to develop a conditional generative framework capable of jointly predicting spatial deformations and post-operative translated outcomes, as illustrated in Figure~\ref{fig:intro-tracetrans}. Conventional registration approaches, such as VoxelMorph \cite{bib:VoxelMorph}, are effective at estimating deformation fields between a fixed and a moving image, yet they require a fixed reference image, which is unavailable in translation tasks where only the source image is provided. Moreover, these methods focus solely on deformation estimation and cannot generate images that align with the target distribution. In contrast, our goal integrates deformation prediction and image synthesis within a unified end-to-end network. This design allows the model to generate anatomically consistent translations while simultaneously providing pixel-level correspondences between the translated and source images, thereby enhancing both interpretability and reliability in clinical applications such as surgical outcome prediction.

\begin{figure}[t!]
    \centering
    \includegraphics[width=\linewidth]{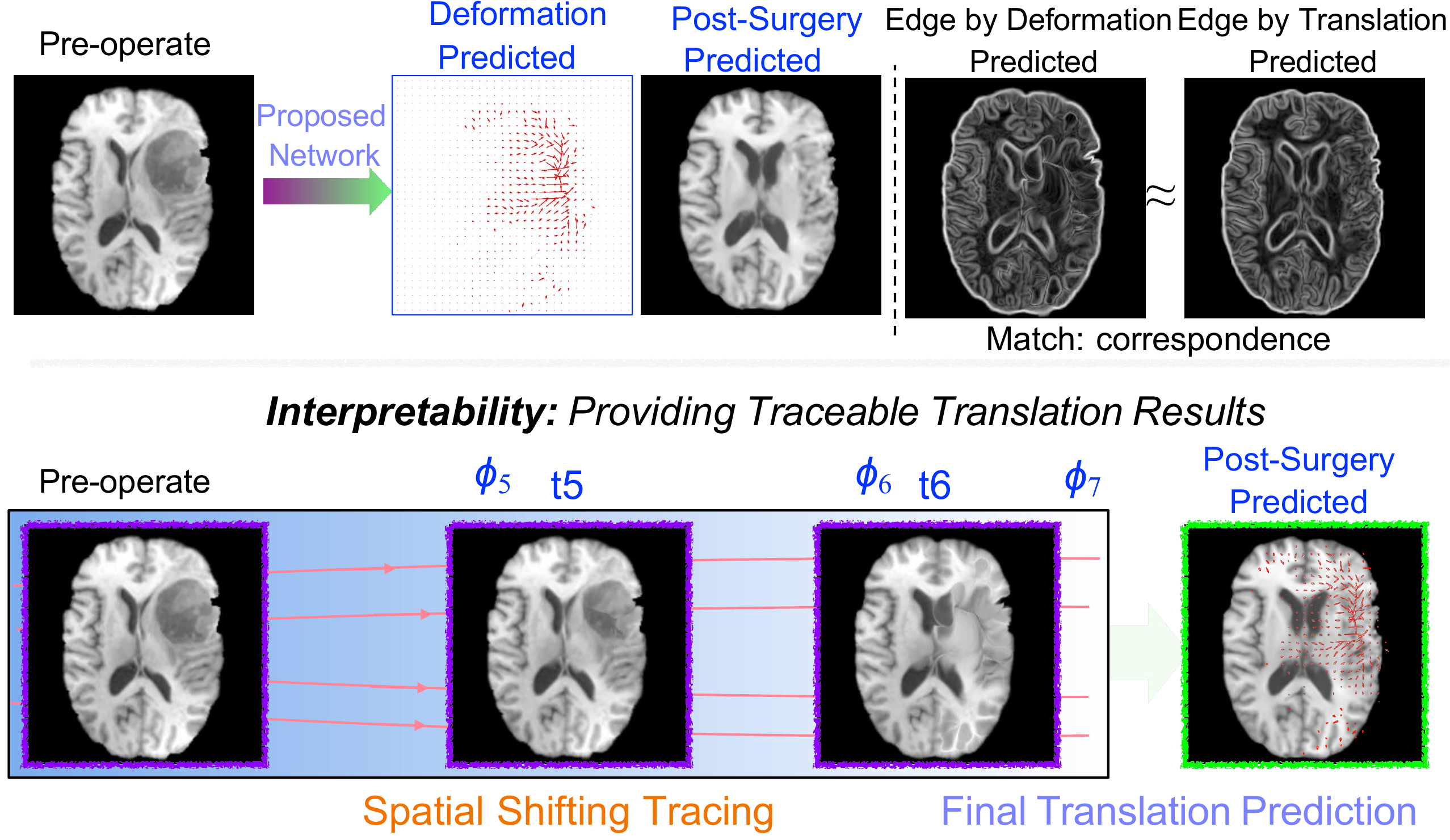}
    \caption{TraceTrans jointly predicts spatial deformations and post-operative translations, enabling pixel-level correspondences and traceable structural changes for interpretable surgical outcome prediction. The second row illustrates the translation process, where spatial shifts are composed from velocity fields, while the first row shows that the close alignment between edges of the deformed pre-operative based on the predicted deformation and predicted post-operative images confirms anatomical correspondence.}
    \label{fig:integration_visualization}
\end{figure}

Therefore, we propose TraceTrans, a novel two-stream end-to-end model for surgical outcome prediction. To the best of our knowledge, this is the first approach to generate predicted post-operative images with spatial traceability without requiring fixed reference images as additional inputs. The model consists of an encoder for feature extraction and two decoders that predict a velocity field and synthesize the translated image. The velocity field is integrated into a deformation field through an iterative composition process, and the resulting spatial changes are illustrated in Figure~\ref{fig:integration_visualization}. To preserve anatomical structure in the translated images, multiple structural constraints are incorporated into the training objective, as detailed in Sec. Method. We evaluate TraceTrans on both medical cosmetology and BraTS-Reg \cite{baheti2021brain} datasets, which contain pre-operative–post-operative face image pairs and brain MRI slice pairs. Quantitative and qualitative comparisons with existing image translation models demonstrate that TraceTrans produces higher-quality translations with superior structural consistency to the source images, as presented in Sec. Experiments.

Our main contributions can be summarized as follows.
\begin{itemize}
    \item We define the interpretable surgical prediction task as generating spatially traceable post-operative images solely from pre-operative inputs.
    \item We propose TraceTrans, a novel two-stream end-to-end model for surgical outcome prediction. To the best of our knowledge, this is the first approach to generate predicted post-operative images with spatial traceability, without requiring fixed reference images during translation training to capture structural changes.
    \item TraceTrans is evaluated on two representative scenarios: predicting facial structural changes after cosmetic surgery and modeling longitudinal brain MRI changes in glioma patients, with results demonstrating superior effectiveness compared to prior approaches.
\end{itemize}

\section{Related Works}
\subsection{Image-to-Image Translation}
Image-to-image translation is a core task in computer vision and aims to generate image in target domain based on image in source domain while preserving the structure of subject in the image. From the perspective of dataset, current image-to-image translation methods can be roughly categorized into supervised methods and unsupervised methods that require datasets with paired data and two datasets in different domains, respectively, and from the perspective of architecture, Generative Adversarial Network (GAN) \cite{bib:GAN} and diffusion model \cite{bib:DDPM} are the most popular basic architectures for this task, and many works are adapted from one of them. 

Pix2Pix \cite{bib:pix2pix} and CycleGAN \cite{bib:cyclegan} are pioneering works in supervised GAN and unsupervised GAN and both lay a solid foundation for the development of GAN-based image-to-image translation models \cite{bib:pix2pixHD, bib:GcGAN, bib:RegGAN, bib:EdgeGuidedGAN, bib:DeformationAwareGan}. Palette \cite{bib:palette} is the first systematic and diffusion-based image translation framework that achieves outstanding performance in multiple image translation tasks. After the success of Palette, many diffusion-based image-to-image translation models are proposed, such as Latent Diffusion Models \cite{bib:latentDiffusion}, DiffI2I \cite{bib:DiffI2I}, DDIB \cite{bib:DDIB} and so on. Although these models are capable of generating good translated images, it is difficult to maintain a precise structure correspondence between the translated image and the source image. Hence, they cannot be directly applied to surgical prediction task.

\subsection{Surgical Prediction}
Prediction of the natural progression of disease and prediction of surgical outcome are two important tasks that facilitate the diagnosis and planning of surgery. With the advancement of image-to-image translation models, many works adapt them, such as GAN-based models \cite{bib:disease-prediction-gan1, bib:disease-prediction-gan2, bib:surgical-prediction1, bib:surgical-prediction2} and diffusion-based models \cite{bib:disease-prediction-diffusion1} to directly synthesize the outcome image and achieve excellent results in their specific field. We choose the GAN model as our backbone due to its high response ability, and our aim is to provide a more robust and reliable surgical outcome prediction model.

\begin{figure*}[t!]
\centering
\includegraphics[width=\textwidth]{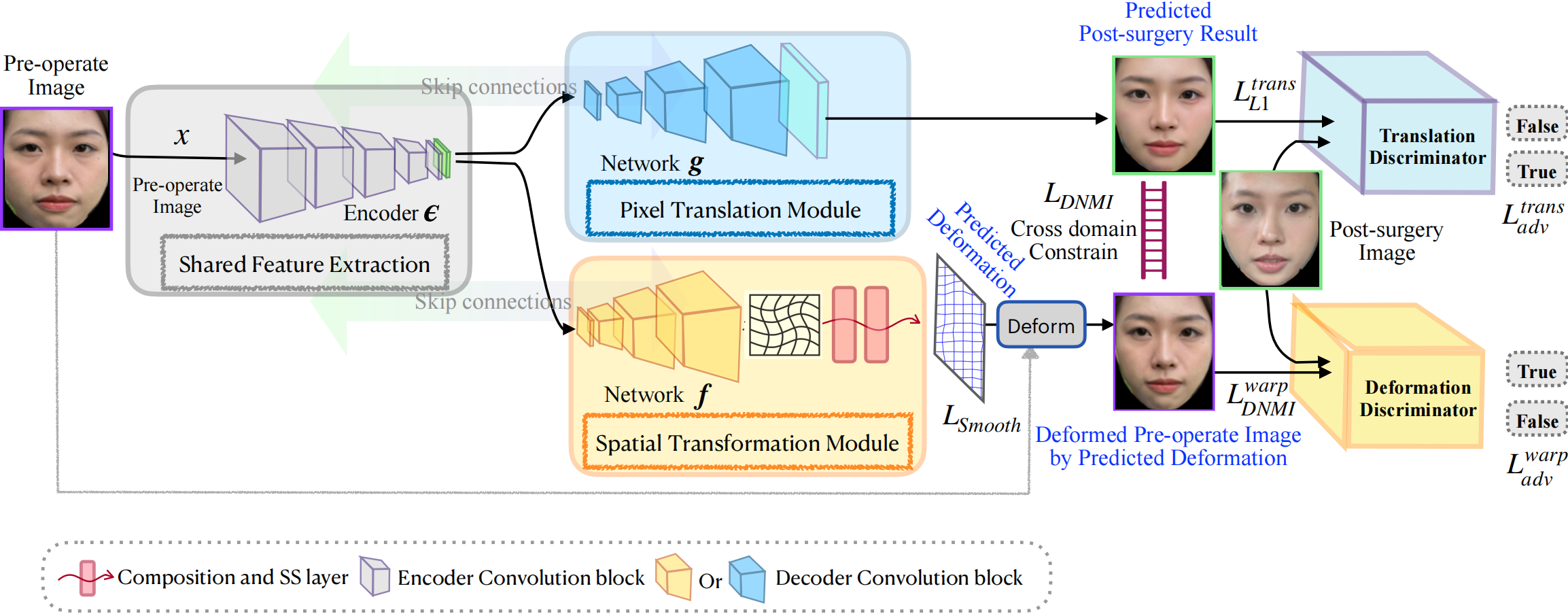} 
\caption{Overview of the TraceTrans framework. An encoder extracts features from the pre-operative image, which are processed by two decoders: the pixel translation module generates the post-operative prediction, and the spatial transformation module predicts the deformation field. The deformation ensures structural correspondence between the pre-operative and post-operative domains, while discriminators and multi-level constraints jointly optimize translation fidelity and anatomical consistency.}
\label{fig:Architecture}
\end{figure*}

\subsection{Medical Image Registration}
Medical image registration aims to establish the spatial correspondence between two images that shares the same structure for the subject and is the crucial and fundamental step in medical image analysis. Traditional methods solve the best registration by optimizing a predefined similarity metric \cite{bib:registration-BSplines, bib:ANTs} and the optimization process is relatively slow. Deep learning-based methods, such as VoxelMorph \cite{bib:VoxelMorph}, accelerate the prediction process by modeling registration as a regression problem. Due to the outstanding performance of VoxelMorph, many follow-up works \cite{bib:registration-dl1, bib:registration-dl2, cheng2025sacb} are adapted from it. Although these methods are capable of predicting the registration with high precision, they cannot be applied directly to surgical prediction, since these models require fixed images as input, and the fixed images are unavailable before prediction.

\section{Method}
\label{sec:method}
The backbone of TraceTrans adopts a dual-stream GAN architecture, consisting primarily of an encoder $\boldsymbol{\epsilon}$, followed by a pixel translation decoder $g$, and a parallel deformation flow decoder $f$.  
Two corresponding discriminators are employed to distinguish between the translated and deformed outputs, respectively.  
An overview of TraceTrans's architecture is illustrated in Figure~\ref{fig:Architecture}.  
We begin by defining the deformation process in Sec. Image Deformation, then describe the structure of the model's backbone in Sec. Model Architecture, and finally provide details of the training procedure in Sec. Training.

\subsection{Image Deformation}

\label{sec:image deformation}
Given $\boldsymbol{m}$ as the $n$-dimensional source (moving) image defined over grid $\Omega \subset \mathbb{R}^n$, let $\mathbf{u}: \Omega \rightarrow \mathbb{R}^n$ and $\mathbf{v}: \Omega \rightarrow \mathbb{R}^n$ be the displacement field and the velocity field, respectively. To encourage a diffeomorphic deformation, we integrate the predicted stationary velocity field $\mathbf{v}$ to form the displacement field $\mathbf{u}$ using scaling and squaring layers \cite{ashburner2007fast}. The displacement field $\mathbf{u}_T$ at any time $T$ is integrated from the velocity field $\mathbf{v}$ according to 
\begin{equation}
    \mathbf{u}_T(\mathbf{p}) = \int_0^T \mathbf{v}((Id + \mathbf{u}_t)(\mathbf{p})) dt
    \text{.}
\label{eq:defrom integration}
\end{equation}
In our code implementation, we adopt $T = 1$ to obtain the final displacement field $\mathbf{u} = \mathbf{u}_1$, using a 7-step integration. Consequently, the resulting deformation field $\boldsymbol{\phi}$ can be expressed as
\begin{equation}
    \boldsymbol{\phi}(\mathbf{p}) = (Id + \mathbf{u})(\mathbf{p})
    \text{,}
\label{eq:deformation field}
\end{equation}
where $\mathbf{p} \in \Omega$ is an arbitrary spatial point of dimension $n$ in the moving image $\boldsymbol{m}$, which is, in the 2D case, a pixel. $Id(\mathbf{p}) = \mathbf{p}$ is an identity deformation. Once the deformation field $\boldsymbol{\phi}$ is predicted, the warped image $\boldsymbol{m}'$ can be generated by $\boldsymbol{m}' = \boldsymbol{m} \circ \boldsymbol{\phi}$ if $\boldsymbol{m}$ is defined over $\mathbb{R}^n$. 

Note that $\boldsymbol{m} \circ \boldsymbol{\phi}(\mathbf{p})$ is not defined for $\boldsymbol{\phi}(\mathbf{p}) \notin \Omega$, interpolation is needed to estimate the value of $\boldsymbol{m}$ at $\boldsymbol{\phi}(\mathbf{p}) \notin \Omega$. Hence the spatial transformer network \cite{bib:SpatialTransformer} is utilized to perform a differentiable linear interpolation at a point $\boldsymbol{\phi}(\mathbf{p}) \notin \Omega$. More specifically, the linear interpolation process is defined as
\begin{equation}
    \boldsymbol{m}'(\mathbf{p}) \approx \sum_{\mathbf{q} \in Z(\mathbf{p}')} \boldsymbol{m}(\mathbf{q}) \prod_{i = 1}^n (1-| \mathbf{q}_i - \mathbf{p}'_i|)
    \text{,}
\end{equation}
where $\boldsymbol{m}'$ is the warped image, $\mathbf{p}' = \boldsymbol{\phi}(\mathbf{p})$, and $Z(\mathbf{p'})$ is the set of $2^n$ discrete neighbor points of $\mathbf{p'}$.

\subsection{Model Architecture}
\label{sec:model arch}
As illustrated in Figure~\ref{fig:Architecture}, the network comprises three components:  
1) an encoder $\boldsymbol{\epsilon}$ hierarchically extracts multi-scale features from the pre-operative images;  
2) two decoders, $g$ and $f$, are employed to predict pixel translations of post-operative images and to estimate spatial transformations, respectively;  
3) two discriminators, $D_{\text{trans}}$ and $D_{\text{warp}}$, are used to distinguish the results of translated post-operative and the deformed pre-operative images, respectively.

Given a pre-operative image $\mathbf{x}$, the encoder $\boldsymbol{\epsilon}$ extracts multi-scale features $\boldsymbol{\epsilon}(\mathbf{x})$.  
Then, the translation decoder $g$ takes these multi-scale features to generate the translated post-operative image as:

\begin{equation}
    \mathbf{y}_{\text{trans}} = g(\boldsymbol \epsilon(\mathbf{x}))
    \text{.}
\end{equation}
In the Spatial Transformation Module, we note that the warped result $\mathbf{y}_{\text{warp}}$ is not directly generated by the deformation  decoder $f$.  
Instead, $f$ produces a velocity field $\mathbf{v}$:
\begin{equation}
    \mathbf{v} = f(\boldsymbol{\epsilon}(\mathbf{x})) \text{.}
\end{equation}
The final deformation field $\boldsymbol{\phi}$ is then obtained by integrating the velocity field $\mathbf{v}$, as defined in Eq.~\ref{eq:defrom integration} and Eq.~\ref{eq:deformation field}.
Next, $\boldsymbol{\phi}$ is applied to warp the pre-operative image $\mathbf{x}$, producing the deformed result:
\begin{equation}
    \mathbf{y}_\text{warp} = \mathbf{x} \circ \boldsymbol{\phi}
    \text{.}
\end{equation}

Suppose the ground truth post-operative reference image is $\mathbf{y}$. The translated $\mathbf{y}_{\text{trans}}$ is combined with $\mathbf{y}$ to form a fake-real pair, and fed to the translation discriminator $D_{\text{trans}}$. Similarly, $\mathbf{y}_{\text{warp}}$ is also combined with $\mathbf{y}$ and fed to the warp discriminator $D_{\text{warp}}$. Our discriminators adopt the conditional GAN formulation, in which the source image $\mathbf{x}$ is used as a condition to guide the discrimination process \cite{bib:cGAN} and are discussed in the next Section.

To ensure structural correspondence between $\mathbf{y}_{\text{trans}}$ and $\mathbf{y}_{\text{warp}}$, a cross-domain constraint is applied between the two images. This constraint is implemented via differentiable normalized mutual information (DNMI) loss.

\subsection{Training}
\label{trainingdetails}
\subsubsection{DNMI}
We adopt differentiable mutual information to compute the loss between generated $\mathbf{y}_\text{warp}$ and reference $\mathbf{y}$, as well as the cross-domain similarity between $\mathbf{y}_\text{trans}$ and $\mathbf{y}_\text{warp}$, in order to include the cross-domain constraint in the final backpropagation. The definition of normalized differentiable mutual information \cite{bib:DNMI} is given as follows.
\begin{equation}
    NMI(X,Y) = \frac{H(X) + H(Y)}{H(X,Y)}
\end{equation}
where $X$, $Y$ are two random variables, $H(X)$ is the entropy. Note that NMI is not differentiable, Parzen window \cite{bib:ParzenWindow} is used to substitute for original rectangular window. Since the more two images are structurally similar to each other, the higher the DNMI is, $\mathcal{L}_{\text{DNMI}}(x,y)$ is formulated as $\mathcal{L}_{\text{DNMI}}(x,y) = -DNMI(x,y)$.

\subsubsection{Trans-Deform Ratio}
\label{sec:trans-deform-ratio}
In order to clearly define the computation of losses of our model, it is necessary to explain the controllable ratio of the two GAN streams in TraceTrans. Since the two decoders $g$ and $f$ are responsible for translation and deformation tasks respectively, we can control the backpropogation proportions of the two tasks.  We define the proportion of the translation task in TraceTrans as $\alpha$, and the deformation task $1 - \alpha$. $\alpha$ is a hyperparameter that can be adjusted during training.

A content alignment loss $\mathcal{L}_{\text{align}}$ is defined to measure the semantic consistency between the generated outputs and their corresponding targets, as well as the cross-domain constraint: 
\begin{equation}
\begin{split}
    \mathcal{L}_{\text{align}} = & 
    \alpha \mathcal{L}_{\text{L1}}^{\text{trans}} + 
    (1 - \alpha) \mathcal{L}_{\text{DNMI}}^{\text{warp}} 
    \\ & + \min(\alpha, 1 - \alpha) \cdot \gamma \cdot \mathcal{L}_{\text{DNMI}}^{\text{cross}}
    \text{,}
\end{split}
\end{equation}
where $\mathcal{L}_{\text{L1}}^{\text{trans}} = \|\mathbf{y}_\text{trans} - \mathbf{y}\|_1$,
$\mathcal{L}_{\text{DNMI}}^{\text{warp}} = -DNMI(\mathbf{y}_\text{warp},\mathbf{y}) $,
and $\mathcal{L}_{\text{DNMI}}^{\text{cross}} = -DNMI(\mathbf{y}_\text{trans},\mathbf{y}_\text{warp}) $.
Here, $\gamma$ is used to control the strength of cross-domain constraint. The sensitivity analysis of $\gamma$ is discussed in Sec: Cross-Domain Constraint.

\subsubsection{Training Framework}
The adversarial loss $\mathcal{L}_{\text{adv, G}}$ for the generator $G$ and $\mathcal{L}_{\text{adv, D}}$ for the discriminator $D$ are defined by the minimax game, 
\begin{equation}
    \begin{split}
    \theta_{G}^* & = \mathop{\arg \min}_{\theta_G}\max_{\theta_D}  \left[ \mathbb{E}_{x\in X}\left\vert D(x, y(x))\right\vert^2 \right. \\ & \left. + \mathbb{E}_{x \in X}\left\vert1-D(x,G(x))\right\vert^2 \right]
    \text{,}
    \end{split}
\end{equation}
and written as follows, respectively.
\begin{equation}
    \mathcal{L}_{\text{adv, G}}(\theta_{G}) = \vert D(x,G(x)) - 1 \vert^2
\end{equation}
\begin{equation}
\begin{split}
    \mathcal{L}_{\text{adv, D}}(\theta_{D}) = \frac{1}{2} \left(\vert D(x ,y(x)) - 1 \vert^2 +  \vert D(x ,G(x))\vert^2 \right)
\end{split}
\end{equation}
where $\theta_{G}$, $\theta_{D}$ denote the parameters for $G$, $D$ respectively; $\theta_G^*$ denotes the best parameter for $G$; $X$ is the set of source images, $y(x)$ is the target image corresponding to the source image $x$.

\begin{table*}[t]
\centering
\setlength{\tabcolsep}{6pt}
\resizebox{0.8\textwidth}{!}{%
\begin{tabular}{l|cccc}
\toprule \textbf{Method} & SSIM(\%)$\uparrow$ & MAE$\downarrow$ & PSNR$\uparrow$ & NMI$\uparrow$ \\
\midrule
MUNIT \cite{bib:munit}          & 79.64~(\textpm 4.11)  & 22.65~(\textpm 10.57)  & 28.96~(\textpm 2.90)  & 1.1670~(\textpm 0.0199) \\
CycleGAN \cite{bib:cyclegan}    & 78.51~(\textpm 4.63)  & 17.92~(\textpm 4.50)  & 29.33~(\textpm 1.93)  & 1.1802~(\textpm 0.0333) \\
GcGAN \cite{bib:GcGAN}          & 77.67~(\textpm 4.55)  & 18.81~(\textpm 5.21)  & 29.26~(\textpm 1.97)  & 1.1738~(\textpm 0.0314) \\
Pix2Pix \cite{bib:pix2pix}      & 79.93~(\textpm 4.20)  & 16.72~(\textpm 3.98)  & 29.60~(\textpm 1.08)  & 1.1821~(\textpm 0.0446) \\
Palette \cite{bib:palette}      & 69.80~(\textpm 4.55)  & 25.38~(\textpm 7.55)  & 27.63~(\textpm 2.25)  & 1.1737~(\textpm 0.0242) \\
DDIB \cite{bib:DDIB}            & 74.63~(\textpm 5.72)  & 20.47~(\textpm 8.53)  & 28.32~(\textpm 2.44)  & 1.1745~(\textpm 0.0243) \\
\textbf{TraceTrans} & \textbf{82.92~(\textpm 4.14)} & \textbf{14.37~(\textpm 4.41)} & \textbf{29.86~(\textpm 0.93)} & \textbf{1.1943~(\textpm 0.0414)} \\
\bottomrule
\end{tabular}
}
\caption{Quantitative comparison on the Face Cosmetology datasets. TraceTrans obtained the best result in all the metrics on this datasets.}
\label{tab:face_results}
\end{table*}
\begin{table*}[t]
\centering
\setlength{\tabcolsep}{6pt}
\resizebox{0.8\textwidth}{!}{%
\begin{tabular}{l|cccc}
\toprule
\textbf{Method} & SSIM(\%)$\uparrow$ & MAE$\downarrow$ & PSNR$\uparrow$ & NMI$\uparrow$ \\
\midrule
MUNIT \cite{bib:munit}          & 72.10~(\textpm 7.35)  & 13.35~(\textpm 4.31)  & \textbf{31.52~(\textpm 2.39)}  & 1.2144~(\textpm 0.0235) \\
CycleGAN \cite{bib:cyclegan}    & 72.10~(\textpm 7.59)  & 14.95~(\textpm 4.59)  & 31.45~(\textpm 2.58)  & 1.2080~(\textpm 0.0218) \\
GcGAN \cite{bib:GcGAN}          & 72.67~(\textpm 8.24)  & 15.06~(\textpm 4.92)  & 31.28~(\textpm 2.72)  & 1.2100 ~(\textpm 0.0240) \\
Pix2Pix \cite{bib:pix2pix}      & 74.01~(\textpm 6.85)  & 14.74~(\textpm 3.91)  & 31.30~(\textpm 0.48)  & 1.2112~(\textpm 0.0291) \\
Palette \cite{bib:palette}      & \: 61.96~(\textpm 10.23)  & 21.19~(\textpm 8.03)  & 29.10~(\textpm 2.48)  & 1.2055~(\textpm 0.0262) \\
DDIB \cite{bib:DDIB}            & 65.47~(\textpm 6.46)  & 18.31~(\textpm 4.26)  & 31.18~(\textpm 1.74)  & 1.1864~(\textpm 0.0146) \\ 
\textbf{TraceTrans} & \textbf{75.97~(\textpm 6.90)} & \textbf{13.30~(\textpm 3.93)} & 31.48~(\textpm 0.44) & \textbf{1.2181~(\textpm 0.0285)} \\
\bottomrule
\end{tabular}
}
\caption{Quantitative comparison on the Brain MRI datasets. TraceTrans obtained the best result in the majority of metrics on this datasets.}
\label{tab:mri_results}
\end{table*}

\begin{figure*}[t!]
    \centering
    \begin{subfigure}[b]{0.11\textwidth}
        \includegraphics[width=\textwidth]{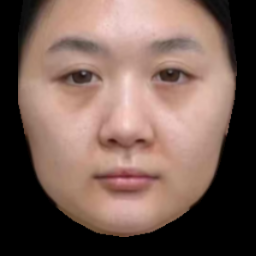}
        \includegraphics[width=\textwidth]{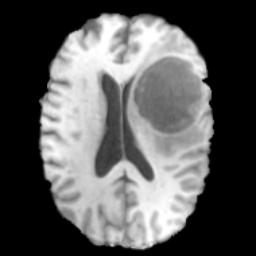}
        \caption*{Pre-operative}
    \end{subfigure}
    \begin{subfigure}[b]{0.11\textwidth}
        \includegraphics[width=\textwidth]{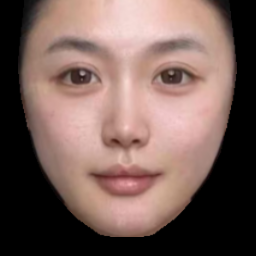}
        \includegraphics[width=\textwidth]{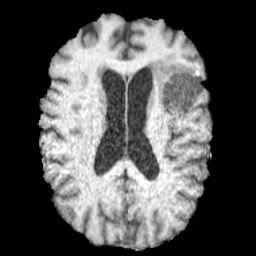}
        \caption*{Reference}
    \end{subfigure}
    \begin{subfigure}[b]{0.11\textwidth}
        \includegraphics[width=\textwidth]{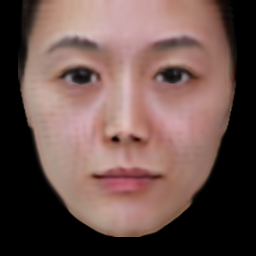}
        \includegraphics[width=\textwidth]{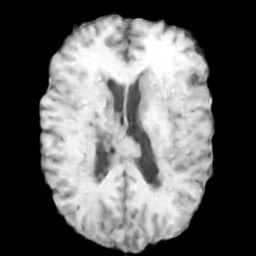}
        \caption*{MUNIT}
    \end{subfigure}
    \begin{subfigure}[b]{0.11\textwidth}
        \includegraphics[width=\textwidth]{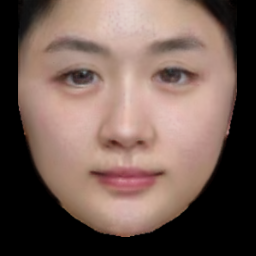}
        \includegraphics[width=\textwidth]{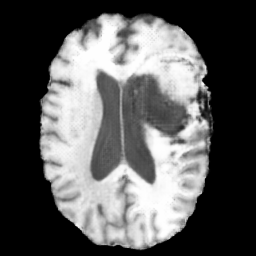}
        \caption*{GcGAN}
    \end{subfigure}
    \begin{subfigure}[b]{0.11\textwidth}
        \includegraphics[width=\textwidth]{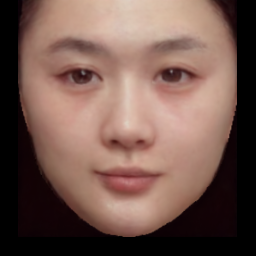}
        \includegraphics[width=\textwidth]{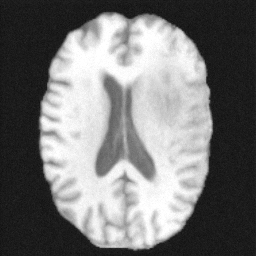}
        \caption*{Palette}
    \end{subfigure}
    \begin{subfigure}[b]{0.11\textwidth}
        \includegraphics[width=\textwidth]{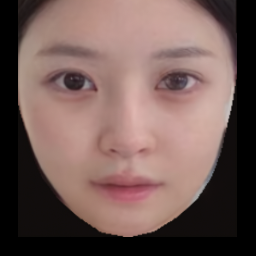}
        \includegraphics[width=\textwidth]{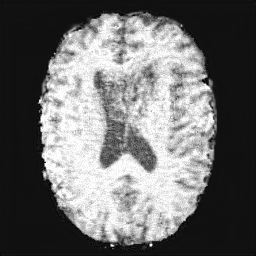}
        \caption*{DDIB}
    \end{subfigure}
    \begin{subfigure}[b]{0.11\textwidth}
        \includegraphics[width=\textwidth]{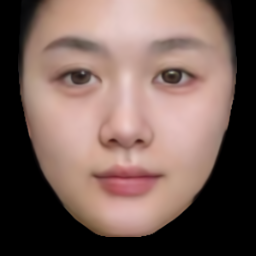}
        \includegraphics[width=\textwidth]{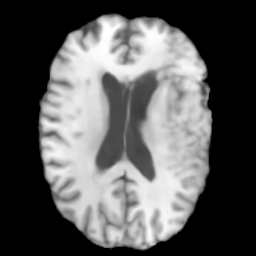}
        \caption*{Ours}
    \end{subfigure}
    \begin{subfigure}[b]{0.11\textwidth}
        \includegraphics[width=\textwidth]{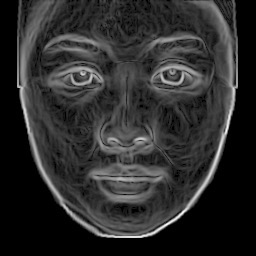}
        \includegraphics[width=\textwidth]{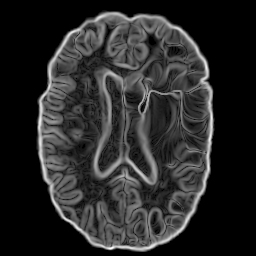}
    \caption*{Deformed edge}
    \end{subfigure}

    \caption{Qualitative comparison of our method and several baselines. The images in the first row and the second row are the results on face cosmetology dataset and brain MRI dataset, respectively. The first two columns show the input and ground truth images, the last column shows Sobel edge maps derived from the deformed source images using the predicted deformation field of our method.}
    \label{fig:qualitative_comparison}
\end{figure*}

\subsubsection{Overall Loss Function }
\label{sec:overall-losses}
Given the trans-proportion $\alpha$ and the content alignment loss $\mathcal{L}_{\text{align}}$, the overall generator loss $\mathcal{L}_G$ is computed as
\begin{equation}
\begin{split}
    \mathcal{L}_G = & 
    \mathcal{L}_{\text{align}} + 
    \lambda_\text{adv}(\alpha\mathcal{L}_{\text{adv, G}}^{\text{trans}} + 
    (1 - \alpha)\mathcal{L}_{\text{adv, G}}^{\text{warp}})
    \\ & + (1 - \alpha)\lambda_\text{smooth} \mathcal{L}_{\text{smooth}}
    \text{,}
\end{split}
\end{equation}
and the overall discriminator loss $\mathcal{L}_D$ is
\begin{equation}
    \mathcal{L}_D = \lambda_{\text{adv}}(\alpha\mathcal{L}_{\text{adv, D}}^{\text{trans}} +
    (1 - \alpha)\mathcal{L}_{\text{adv, D}}^{\text{warp}})
    \text{.}
\end{equation}

The $\mathcal{L}_{\text{adv, G}}^{\text{trans}}$ and $\mathcal{L}_{\text{adv, G}}^{\text{warp}}$ represent the adversarial losses for the two generation tasks, respectively. Likewise $\mathcal{L}_{\text{adv, D}}^{\text{trans}}$ and $\mathcal{L}_{\text{adv, D}}^{\text{warp}}$ are adversarial losses for two discriminators. The smoothness term $\mathcal{L}_{\text{smooth}}(\mathbf{v}) = \sum_{\mathbf{p} \in \Omega} \Vert \nabla \mathbf{v}(\mathbf{p}) \Vert ^2$ is used to regularize the spatial gradients of the deformation field and to ensure the diffeomorphism of the velocity field.

\section{Experiments}
\label{sec:experiment}
We conduct several experiments and an ablation study to evaluate the effectiveness of our method in surgical prediction, and a sensitivity study to find the best value for hyperparameters $\alpha$ and $\gamma$. Two different domains are considered: face cosmetic surgery and brain tumor surgery. We compare our model with several representative models including both GAN-based and diffusion-based models.

\subsection{Datasets}
\subsubsection{Face Cosmetology}
This dataset is a proprietary collection curated by us and consists of 412 paired frontal face images captured before and after cosmetic surgery. Each pair corresponds to the same individual and was collected with informed consent from the image providers. To ensure consistency across samples, an automated face detection pipeline is applied to locate facial regions in each image, followed by masking of non-facial areas such as background and hair. The cropped facial region is then resized to a standardized resolution of $256 \times 256 \times 3$, where the three channels correspond to RGB color information. After preprocessing, 330 image pairs are used for training and the remaining 82 pairs are reserved for testing.

\subsubsection{Brain MRI}
We also evaluate our method on the BraTS-Reg \cite{baheti2021brain} dataset, which provides pre-operative and follow-up MRI scans for patients with diffuse glioma. The dataset includes multimodal MRI sequences (T1, T2, FLAIR, and T1 contrast-enhanced) collected longitudinally for each subject. To mitigate inter-scan variability caused by different acquisition protocols, histogram standardization \cite{histo_stand} is first applied to the pre-operative and follow-up T1 and T2 scans. We then perform Z-normalization, rescale the intensity to the $[0,1]$ range, and crop the scans to a resolution of $192 \times 192 \times 128$. The pre-operative and follow-up scans are rigidly aligned using the SimpleITK registration framework \cite{Yaniv2018}. For experiments, 160 subjects in BraTS-Reg are split into training and testing sets with a 7:3 ratio and the central eleven slices are extracted to form a dataset with 1232 training pairs and 528 testing pairs.

\subsection{Implementation Details}
Our model is implemented in PyTorch and trained for 500 epochs using a batch size of 8. All experiments are conducted on an NVIDIA A100 GPU with 40GB of VRAM. Input images are resized to $256 \times 256$ and normalized to the range $[-1, 1]$.The network is optimized using the Adam optimizer with a learning rate of $2\times10^{-4}$. The encoder contains 5 downsampling layers, where the number of output channels for each layer is [64, 128, 256, 512, 512]. We use the PatchGAN discriminators in our network as proposed in Pix2Pix \cite{bib:pix2pix}. The optimal hyperparameters used for training are $\alpha = 0.5$, $\gamma = 1.0$, $\lambda_\text{adv} = 0.01$, and $\lambda_\text{smooth} = 0.2$.  
We use 16 bins for $\mathcal{L}_{\text{DNMI}}^{\text{warp}}$ and 32 bins for $\mathcal{L}_{\text{DNMI}}^{\text{cross}}$.

\subsection{Evaluation Metrics}
We evaluate the performance of TraceTrans and baseline methods using four standard image similarity metrics:  
Structural Similarity Index (SSIM)~\cite{bib:ssim}, Peak Signal-to-Noise Ratio (PSNR), Mean Absolute Error (MAE), and Normalized Mutual Information (NMI)~\cite{bib:nmi}.  
Higher values of SSIM, PSNR, and NMI, and a lower value of MAE, indicate better image quality and closer alignment with the ground truth.

\subsection{Experiment Results on Face Cosmetology}
Performance evaluations on the face cosmetology dataset are presented in Table~\ref{tab:face_results}, where TraceTrans outperforms all other models across all metrics.  
The significant improvements in SSIM and NMI indicate TraceTrans's superior ability to preserve structural correspondence.  
In addition, the increase in PSNR and reduction in MAE suggest that the predicted post-operative images generated by TraceTrans exhibit better texture translation quality compared to other models. The qualitative results are shown in the first row of Figure~\ref{fig:qualitative_comparison}, where the predicted post-operative image produced by TraceTrans is visually more similar to the reference image than those generated by the other methods, aligning well with the quantitative results.

\subsection{Experiment Results on Brain MRI}
To further demonstrate the effectiveness of TraceTrans, we additionally evaluated the model on Brain MRI dataset. The quantitative results are shown in Table \ref{tab:mri_results} and TraceTrans performs best in SSIM, MAE, NMI and slightly worse than MUNIT in PSNR. Since the difference between TraceTrans and MUNIT on PSNR is less than 5\% of either standard deviation, then this difference is negligible. TraceTrans shows significant SSIM improvement while matching MUNIT's PSNR and achieving the lowest MAE, demonstrating balanced excellence in both structural and textural translation. Qualitative results are illustrated in the second row of Figure \ref{fig:qualitative_comparison}, the image generated by our model is visually best similar to the reference image.

\subsection{Ablation Study}
To study the necessity of each component, we conducted two ablation studies: one evaluating structural correspondence and the other assessing fidelity of the predicted results, with further details provided in the sensitivity analysis. 
\begin{figure}[h!]
    \centering
    \includegraphics[width=\linewidth]{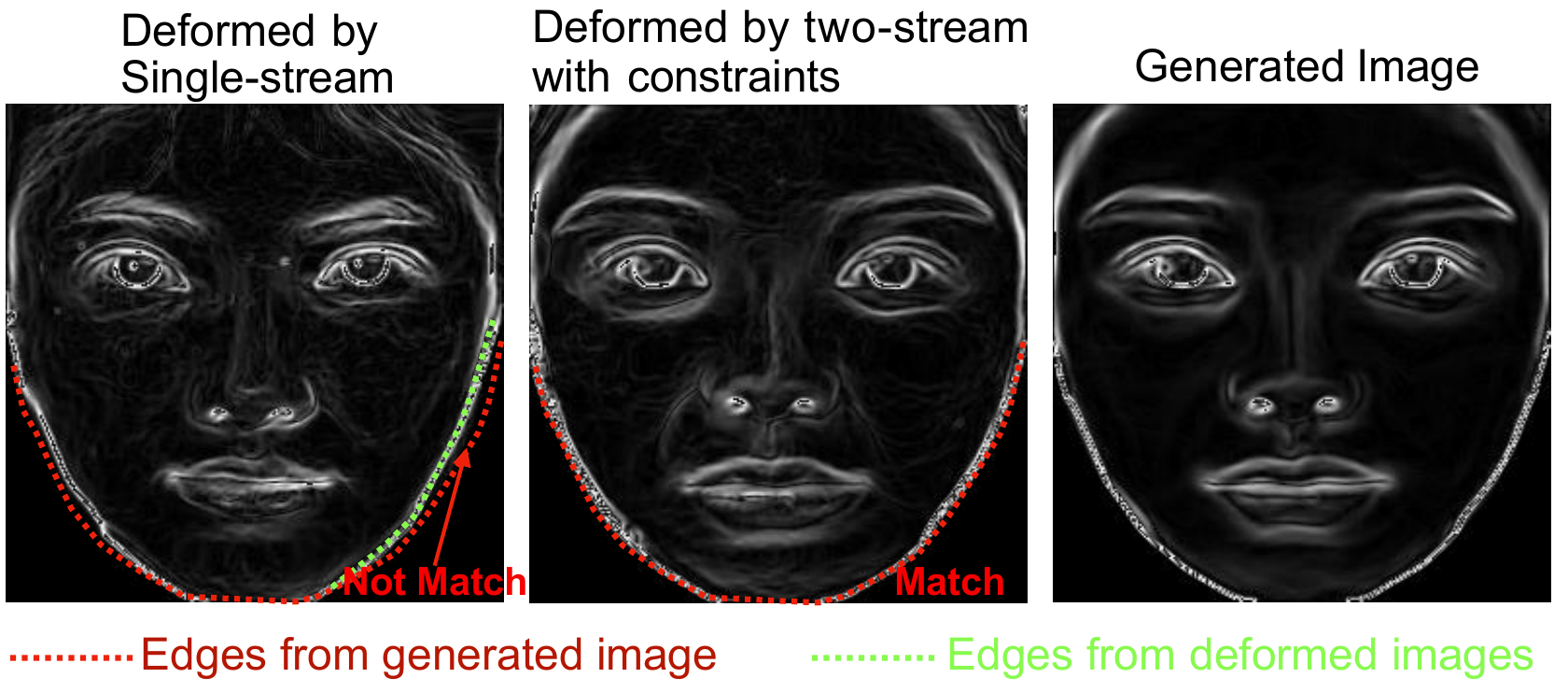}
    \caption{Illustration of correspondence comparison in the ablation study. The proposed two‑stream constrained network generates deformations that align with the translated images, whereas a similar single‑stream architecture fails to preserve this correspondence.}
    \label{fig:placeholder}
\end{figure}

\begin{table}[h!]
\centering
\setlength{\tabcolsep}{2pt} 
\renewcommand{\arraystretch}{1.1} 
\begin{adjustbox}{width=\linewidth}
\begin{tabular}{lccc}
\toprule
\textbf{Correspondence (evaluated by edge)} & SSIM (\%)$\uparrow$ & PSNR$\uparrow$ & NMI$\uparrow$ \\
\midrule
Single-stream (two networks) & 55.58 & 18.24 & 0.2167 \\
Two-stream no constraints & 58.50 & 19.21 & 0.2224 \\
TraceTrans & 59.48 & 19.34 & 0.2249 \\
\bottomrule
\end{tabular}
\end{adjustbox}
\caption{Quantitative comparison of correspondence performance for different model components.}
\label{tab:quantitative_results}
\end{table}
\begin{figure*}[t!]
\centering
  \begin{subfigure}[t]{0.14\textwidth}
    \centering
    \includegraphics[width=\textwidth]{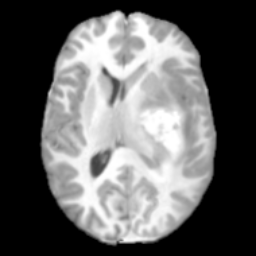}
    \caption{Pre-operative}
    \label{fig:mri_trend_source}
  \end{subfigure}
  \hfill
  \centering
  \begin{subfigure}[t]{0.14\textwidth}
    \centering
    \includegraphics[width=\textwidth]{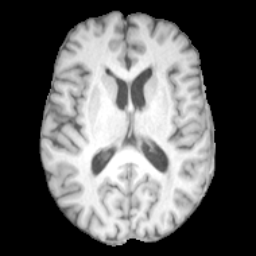}
    \caption{Reference}
    \label{fig:mri_trend_target}
  \end{subfigure}
  \hfill
  \centering
  \begin{subfigure}[t]{0.14\textwidth}
    \centering
    \includegraphics[width=\textwidth]{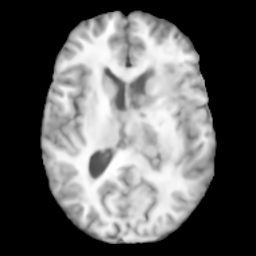}
    \caption{$\alpha = 1$}
    \label{fig:mri_trend_alpha1}
  \end{subfigure}
  \hfill
  \begin{subfigure}[t]{0.14\textwidth}
    \centering
    \includegraphics[width=\textwidth]{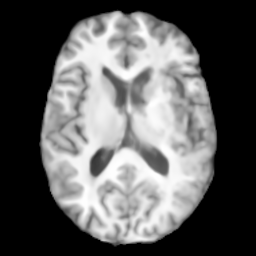}
    \caption{$\alpha = 0.75$}
    \label{fig:mri_trend_alpha0.75}
  \end{subfigure}
  \hfill
  \begin{subfigure}[t]{0.14\textwidth}
    \centering
    \includegraphics[width=\textwidth]{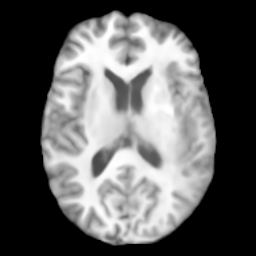}
    \caption{$\alpha = 0.5$}
    \label{fig:mri_trend_alpha0.5}
  \end{subfigure}
  \hfill
  \begin{subfigure}[t]{0.14\textwidth}
    \centering
    \includegraphics[width=\textwidth]{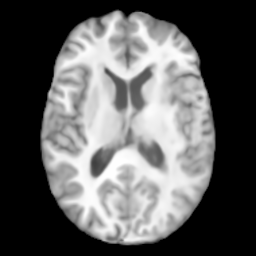}
    \caption{$\alpha = 0.25$}
    \label{fig:mri_trend_alpha0.25}
  \end{subfigure}
  \caption{Brain MRI predictions under $\alpha = 1$, $0.75$, $0.5$ and $0.25$.}
  \label{fig:mri_trend}
\end{figure*}
The evaluation of structural correspondence, which is the most critical aspect, is performed using deformations predicted solely from preoperative images. Specifically, the source image is warped using the predicted deformation field, and Sobel edge maps are computed from the warped image. These are then compared with the Sobel edge maps of the predicted postoperative images to quantify the difference between corresponding edges. For this ablation, three configurations are examined: (1) a single‑stream network that generates postoperative images and deformations in parallel without a shared encoder; (2) a two‑stream network with a shared encoder but without the structural correspondence constraint; and (3) the proposed full model.

Table \ref{tab:quantitative_results} compares correspondence performance across model variants. The single‑stream design without shared encoding or correspondence constraints yields the lowest scores, while adding a two‑stream structure improves all metrics. The full TraceTrans model, which also includes the correspondence constraint, achieves the best results, confirming its effectiveness in preserving edge‑level alignment between preoperative and postoperative predictions.

Additionally, an ablation on image fidelity (Table \ref{tab:ablation}) shows that adding the deformation stream improves MAE, and further introducing the cross-domain constraint achieves the best fidelity, confirming the benefit of the full TraceTrans design.

\begin{table}[ht]
\centering
\begin{adjustbox}{width=\linewidth}
\begin{tabular}{ccc|c}
\toprule
Trans Stream & Deform Stream & Cross-Domain Constraint & Face MAE$\downarrow$ \\
\midrule
\ding{51} & \ding{55} & \ding{55} & 12.88 \\ 
\ding{51} & \ding{51} & \ding{55} & 12.03 \\ 
\ding{51} & \ding{51} & \ding{51} & \textbf{10.64} \\ 
\bottomrule
\end{tabular}
\end{adjustbox}
\caption{Ablation study on image fidelity.}
\label{tab:ablation}
\end{table}

\subsection{Sensitivity Analysis}
\subsubsection{Trans-Defrom Ratio}
\label{sec:alpha}
$\alpha$ controls the contribution of the two streams in our model.  
We evaluate its impact on two datasets, focusing on (1) traceability from input to predicted output and (2) the quality of the prediction.

To assess traceability, we use ANTs~\cite{bib:ANTs} to compute a deformation field from the pre-operative to the predicted post-operative image.  
We then measure similarity between the predicted image and the warped pre-operative image.  
Since the deformation stream is disabled when $\alpha = 1$, its field is excluded for fairness.  
At $\alpha = 1$, the model reduces to Pix2Pix, which often produces blurry MRI outputs, making ANTs registration unstable—though it remains stable on the face cosmetology dataset.

Although both models perform similarly in global facial regions, Pix2Pix struggles to reconstruct key features (e.g., eyes, nose, mouth).  
We address this by masking non-essential areas and computing edge Dice scores within the masked regions.  
Results are reported in the first row of Table~\ref{tab:alpha}. These findings show that the deformation stream enhances both spatial traceability and prediction quality.

\begin{table}[ht]
\centering
\begin{adjustbox}{width=\linewidth}
\begin{tabular}{c|ccccc}
\toprule
$\alpha$ & 1 & 0.75 & 0.5 & 0.25 \\
\midrule
Masked Dice(\%) (face) & 44.72 & 49.26 & 51.10 & 49.62 \\
SSIM(\%) (MRI)         & 74.01 & 75.11 & 75.97 & 75.57 \\
\bottomrule
\end{tabular}
\end{adjustbox}
\caption{Effect of different ($\alpha$) values on traceability and prediction quality. Traceability is measured by Masked Dice between warped input (using ANTs) and predicted output.}
\label{tab:alpha}
\end{table}

We assess output quality by computing the SSIM between predicted and ground-truth post-operative brain MRIs.  
As shown in Table~\ref{tab:alpha}, $\alpha = 0.5$ provides the best trade-off between pixel-level similarity and structural consistency.  
Figure~\ref{fig:mri_trend} presents qualitative comparisons across different $\alpha$ values.

\subsubsection{Cross-Domain Constraint}
\label{sec:gamma}
$\gamma$ is a key parameter that controls the strength of the cross-domain constraint between $\mathbf{y}_\text{trans}$ and $\mathbf{y}_\text{warp}$.  
Although both decoders share the same encoder, it is important to assess whether this constraint is necessary to ensure structural consistency between their outputs.  
We conduct a sensitivity analysis on five $\gamma$ values ranging from 0 to 1 using the face cosmetology dataset.

The cross-domain constraint primarily enforces structural alignment between $\mathbf{y}_\text{trans}$ and $\mathbf{y}_\text{warp}$, which we evaluate using SSIM.  
Within the tested range, larger $\gamma$ values consistently improve structural consistency.  
Additionally, we observe that stronger constraints enhance traceability from the input to the predicted output, as reflected by higher masked Dice scores.  
Results are summarized in Table~\ref{tab:gamma}.

\begin{table}[ht]
\centering
\begin{adjustbox}{width=\linewidth}
\begin{tabular}{c|ccccc}
\toprule
$\gamma$ & 0 & 0.25 & 0.5 & 0.75 & 1.0 \\
\midrule
Masked SSIM(\%) & 83.00 & 83.19 & 83.69 & 85.03 & 85.16 \\
Masked Dice(\%) & 45.60 & 45.70 & 46.79 & 48.96 & 51.10 \\
\bottomrule
\end{tabular}
\end{adjustbox}
\caption{Effect of Cross-Domain Constraint Strength ($\gamma$) on structural correspondence and traceability. Structural correspondence between $\mathbf{y}_\text{trans}$ and $\mathbf{y}_\text{warp}$ is measured by SSIM between unmasked facial areas of these two. Masked Dice is calculated in the same way as Table \ref{tab:alpha}.}
\label{tab:gamma}
\end{table}


 \section{Conclusion}
In this work, we introduced TraceTrans, a deformable image-to-image translation framework for postoperative prediction. Unlike prior methods that focus on distribution alignment and neglect spatial correspondence, TraceTrans  predicts deformation fields to trace structural changes and ensure anatomical consistency. By integrating deformation prediction and image synthesis in an end-to-end design, it produces translations that are both visually accurate and interpretable. Experiments on medical cosmetology and brain MRI datasets show superior translation quality and structural fidelity, demonstrating its potential for clinical applications requiring precise postoperative modeling.
\bibliography{aaai2026}

\onecolumn

\appendix
\setcounter{secnumdepth}{3}

\titleformat{\section}
  {\raggedright\Large\mdseries\scshape}
  {\thesection} 
  {1em}
  {}

\captionsetup{
    font=bf,
    labelfont=normal
}

\section*{\huge{Appendix}}
\vspace{1.5em}

\section{Visualization of Deformation}
\begin{figure}[H]
    \centering
    \includegraphics[width=0.95\linewidth]{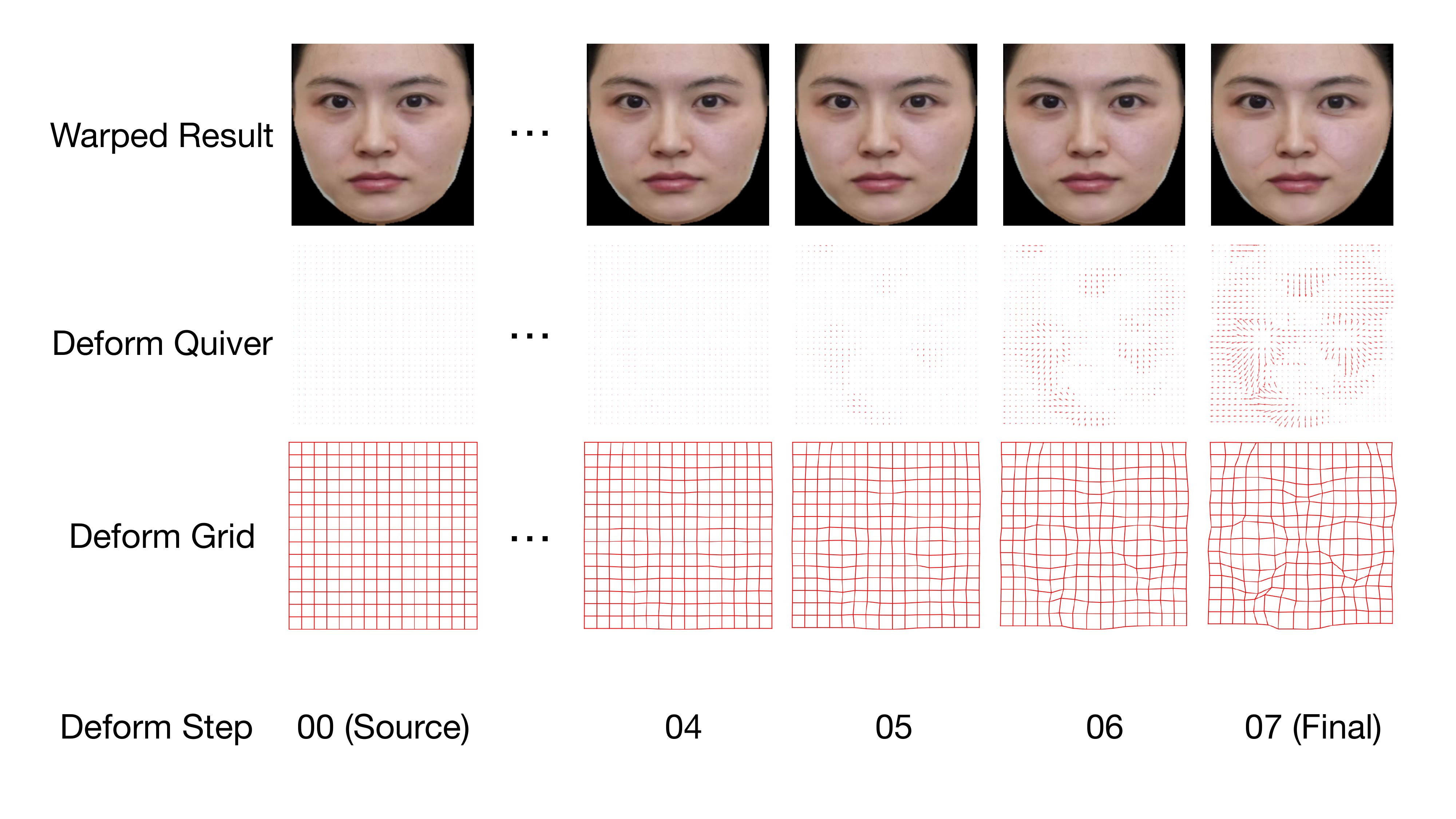}
    \vspace{-0.8em}
    \caption{Deformation Process}
\end{figure}
\vspace{-3em}
\begin{figure}[H]
    \centering
    \includegraphics[width=0.5\linewidth]{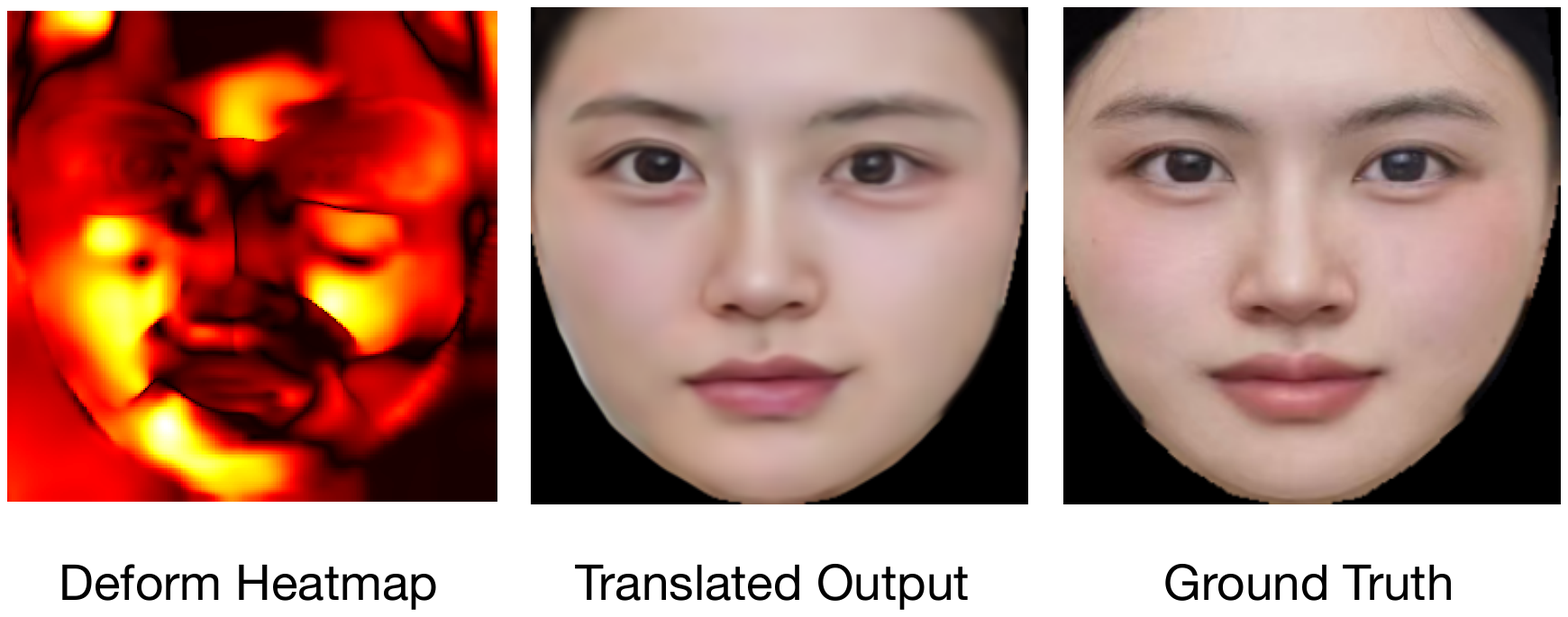}
    \vspace{-3em}
    \caption{Final Translated Output}
\end{figure}

\vspace{2em}

\section{Sobel Edge Comparisons}
Here we show more examples of comparisons between the Sobel edge maps of $\mathbf{y}_{\text{trans}}$ and $\mathbf{y}_{\text{warp}}$.
\begin{figure}[H]
    \centering
    \includegraphics[width=0.95\linewidth]{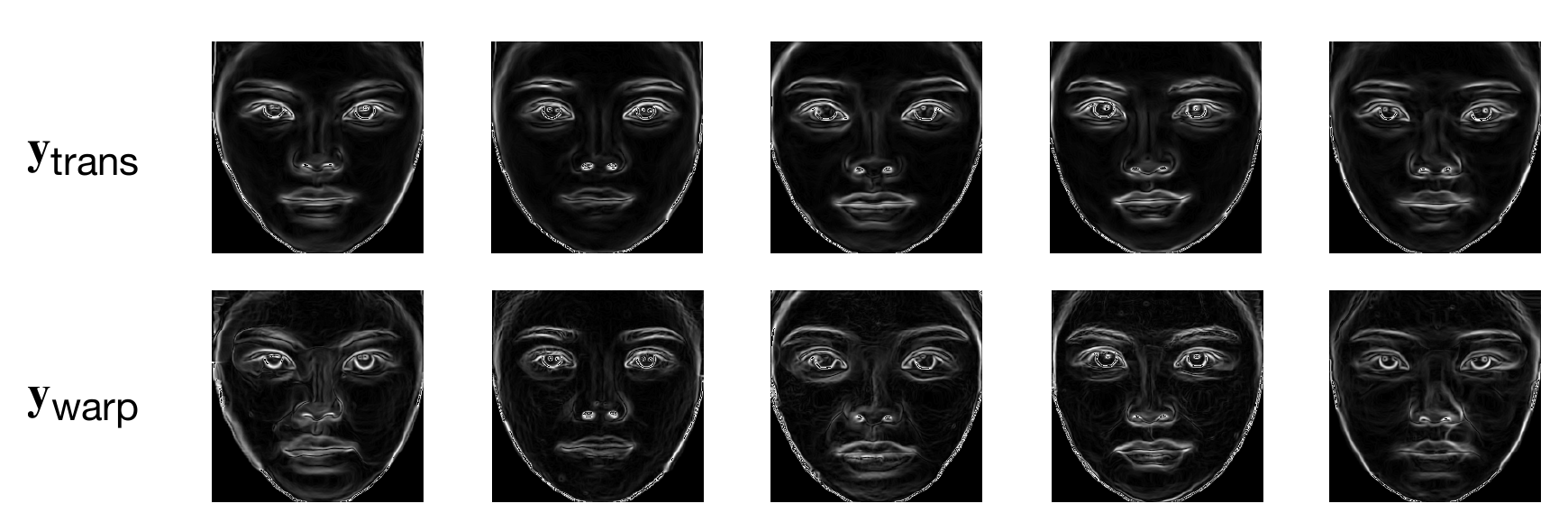}
    \caption{Sobel Edges of Faces}
\end{figure}
\vspace{-1em}
\begin{figure}[H]
    \centering
    \includegraphics[width=0.95\linewidth]{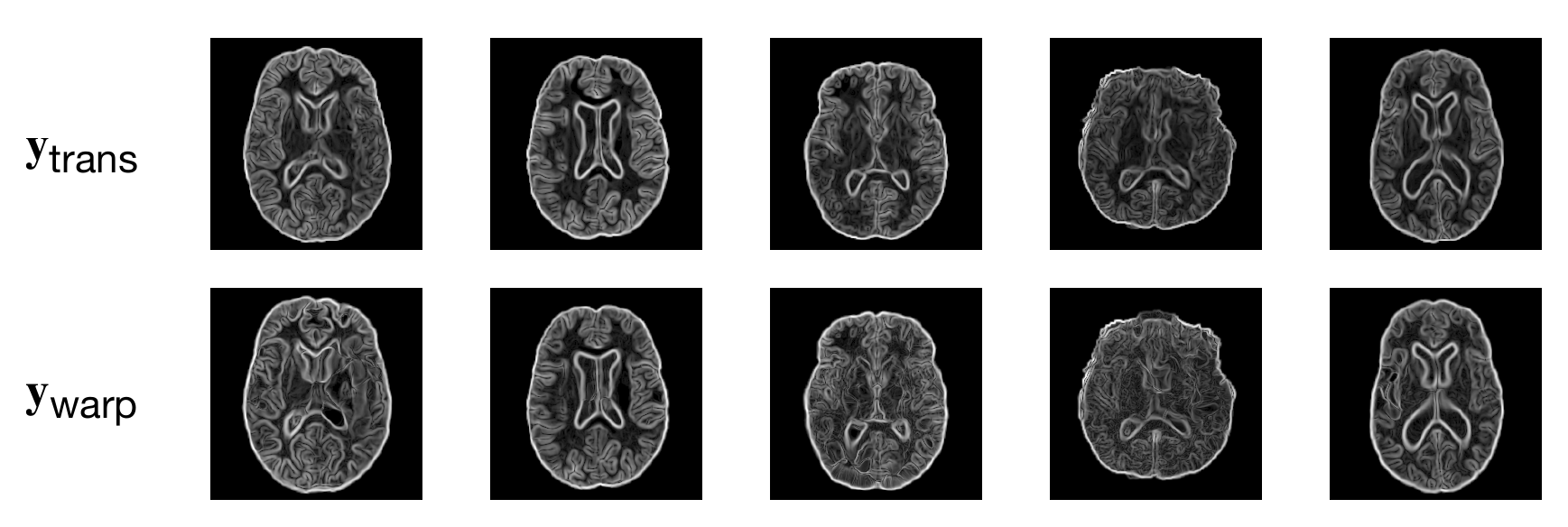}
    \caption{Sobel Edges of Brain MRIs}
\end{figure}

\vspace{2em}

\section{Charts of Sensitivity Analysis}
\begin{figure}[H]
    \centering
    \includegraphics[width=0.9\linewidth]{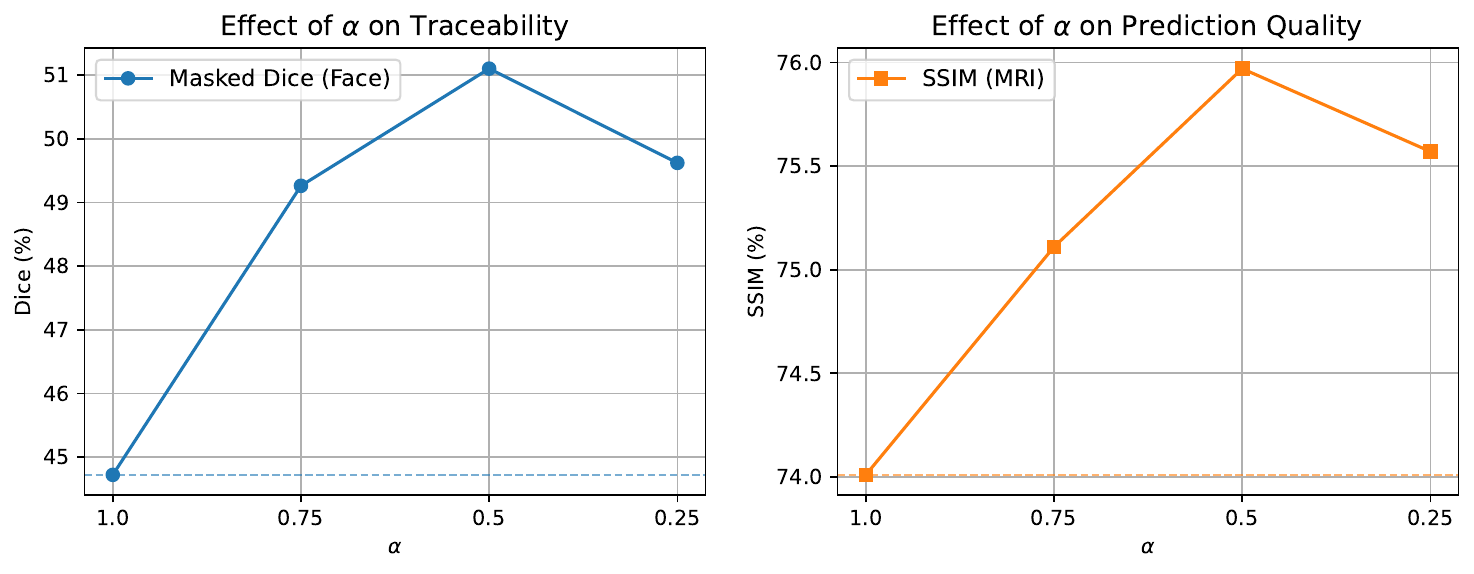}
    \caption{Sensitivity Curve of $\alpha$}
\end{figure}
\begin{figure}[H]
    \centering
    \includegraphics[width=0.9\linewidth]{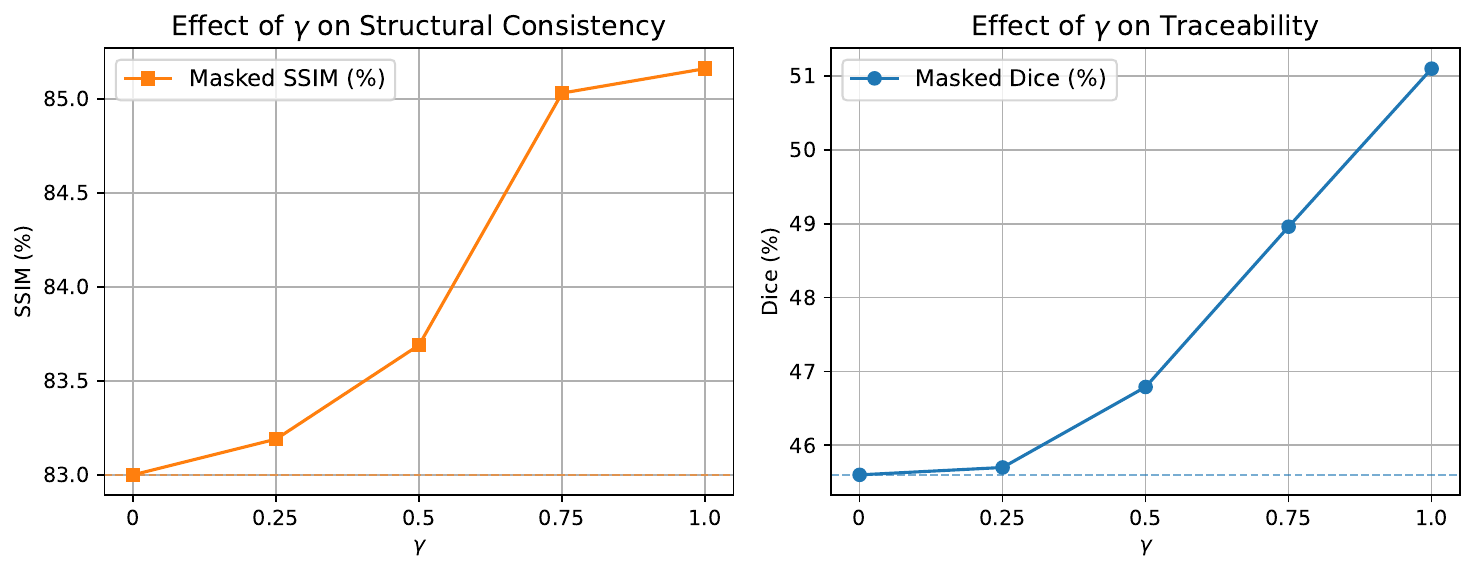}
    \caption{Sensitivity Curve of $\gamma$}
\end{figure}

\end{document}